\documentclass{LT23auth}
\usepackage{graphicx}
\usepackage{bm} 

\def\e{\varepsilon}

\def\gs{{\gamma_{\rm s}}}
\def\ts{{\tau_{\rm s}}}

\def\os{\omega_{S}}

\def\f0{\Phi_0}

\def\gd{\gamma}
\def\muB{\mu_{\rm B}}

\def\cal{\textsl}

\begin{document}

\begin{frontmatter}

\title{Electron energy and phase relaxation on magnetic impurities}

\author{M.G. Vavilov$^a$, A. Kaminski$^{a,b}$ and L.I. Glazman$^a$\thanksref{thank1}}

\address{$^a$Theoretical Physics Institute, University
of Minnesota, Minneapolis, MN 55455, USA
\\
$^b$Condensed Matter Theory Center, Department of
Physics, University of Maryland, College Park, MD 20742 
}

\thanks[thank1]{Corresponding author.  Theoretical Physics Institute,
University of Minnesota, Minneapolis, MN 55455, USA.  E-mail:
glazman@umn.edu}


\begin{abstract}
We discuss the effect of magnetic impurities on the inelastic
scattering and dephasing of electrons. Magnetic impurities mediate the energy exchange between
electrons. This mechanism is especially effective at small energy
transfers $E$ in the absence of Zeeman
splitting, when the two-particle collision integral in the electron
kinetic equation has a kernel $K\propto 1/E^2$ in a broad energy
range. In a magnetic field, this mechanism is suppressed at $E$ below
the Zeeman energy. Simultaneously, the Zeeman splitting of the impurity spin states
reduces the electron dephasing rate, thus enhancing the effect of
electron interference on conduction. We find the weak localization
correction to the conductivity and the magnitude of the conductance
fluctuations in the presence of magnetic field of arbitrary
strength. Our results can be compared quantitatively with the
experiments on energy relaxation in short metallic wires and on
Aharonov-Bohm conductance oscillations in wire rings.
\end{abstract}
\end{frontmatter}

\section{Introduction}

The effect of magnetic impurities on the electron properties of a
metal is drastically different from that of ``usual'' defects which
just violate the translational invariance of the crystalline lattice.
The reason for the difference is that a magnetic impurity brings an
additional degree of freedom -- its spin. We demonstrate that magnetic
impurities may mediate energy transfer between electrons.  If the
transferred energy $E$ exceeds the Kondo temperature $T_K$, then the
energy relaxation occurs predominantly in two-electron collisions.  We
derive the kernel $K$ of the collision integral in the kinetic
equation for the distribution function. This kernel depends strongly
on the transferred energy, $K\propto J^4/E^2$. The dependence of $K$
on the energies $\varepsilon_i$ of the colliding electrons comes from the logarithmic in $|\varepsilon_i|$ renormalization of the exchange integral $J$,
known from the theory of Kondo effect\cite{Hewson},  and is relatively weak as long as $|\varepsilon_i|\gg T_K$. The $1/E^2$
divergence of the kernel is cut off at small $E$; the cut-off energy
is determined by the dynamics of the impurity spins.

Localized spins affect not only the energy relaxation rate, but also
the conventional electron transport properties, such as the
temperature and field dependence of the conductance. No spin dynamics of impurities is
needed for the suppression of the interference corrections to the
conductivity; interaction of electron spins with the magnetic moments
``frozen'' in random directions already leads to that
suppression~\cite{Altshuler2}. Mesoscopic conductance fluctuations
are not suppressed by ``frozen'' magnetic moments.
However, even a relatively slow relaxation (such as Korringa relaxation) 
of individual magnetic moments leads to the time-averaging of the random potential ``seen'' by transport electrons
in the course of measurement, and the mesoscopic fluctuations of the
dc conductance are averaged out.\cite{Bobkovetal} We find the weak localization correction to the conductivity and the magnitude of  conductance fluctuations in the presence of magnetic field of arbitrary strength.

\section{Inelastic scattering of an electron off a magnetic impurity}
We describe a metal with magnetic impurities by means of the
Hamiltonian $\hat H=\hat H_0+\hat V$:
\begin{equation}
\hat{H}_0=\sum_{{\bf k}\alpha}\!\xi^{\phantom{\dagger}}_{\bf k} 
c^\dagger_{{\bf k}\alpha} c^{\phantom{\dagger}}_{{\bf k}\alpha},\ \
\hat V =
J \sum_{\alpha\alpha',l}\! 
\hat{\bf S}_l \bm{\sigma}_{\alpha\alpha'}
\psi^\dagger_{{\mathbf{r}_l}\alpha} 
\psi_{{\mathbf{r}_l}\alpha'},
\label{H}
\end{equation}
where $\hat{\bf S}_l$ is
the spin operator of the $l$-th impurity at point ${\bf r}_l$, 
$\hat{\bf S}^2_l=S(S+1)$. Free electron states 
$c^{\phantom{\dagger}}_{{\bf k}\alpha}$ are labelled
by the wave vector ${\bf k}$ and the spin index $\alpha$, $\psi_{{\mathbf{r}_l}\alpha}=
\sum_{\bf k}e^{i{\bf k}{\mathbf{r}_l}}c^{\phantom{\dagger}}_{{\bf k}\alpha}$. The Pauli matrices are denoted by $\bm{\sigma}\equiv
(\sigma^x,\sigma^y,\sigma^z)$.

The impurities can be considered independently if their concentration $n$   
is low enough. 
In the one-impurity scattering problem, there is interaction only in $s$ channel, so we will label the participating electron states with scalar index $k$.

The lowest
non-vanishing order of the perturbation theory series in the exchange
constant $J$ for the inelastic scattering amplitude is the second
order:
\begin{eqnarray} 
\label{inamp} 
A_{\varrho_1\varrho_2;\varrho_3\varrho_4}^{SS'}= 
\langle \varrho_3\varrho_4,S'| 
\hat{V}\frac{1}{\xi_{k_1}\!\!+\!\xi_{k_2}-\hat{H}_0}
\hat{V} |\varrho_1\varrho_2,S\rangle,
\end{eqnarray}   
where $\varrho_i=(k_i,\alpha_i)$. 
The denominator in Eq.~(\ref{inamp}) is the energy of the intermediate
virtual state, which equals $\pm(\xi_{k_1}-\xi_{k_3})$ for two of the
four possible pairings of the electron creation-annihilation operators,
or $\pm(\xi_{k_1}-\xi_{k_4})$ for the other two pairings. The spin
structure of the scattering amplitude can easily be found from
Eq.~(\ref{inamp}). In a scattering event, spins of one or both
participating electrons must flip, with the corresponding change of
the impurity spin. Here we are interested only in the relaxation of
the electron energy distribution, and assume that in the absence of
magnetic field the system does not have any spin polarization.
Therefore we need to calculate only the total cross-section of
scattering into all possible spin states, averaged over the initial
spin states of the impurity and two electrons. We obtain the collision
integral kernel
\begin{equation}
  \label{KE}
K(E)= \frac{\pi}{2}\frac{n}{\nu}S(S+1)(J\nu)^4
\frac{1}{E^2},
\end{equation}
which depends only on the energy $E$ transferred in the
collision. Here $\nu$ is the electron density of states at the Fermi energy per spin degree of freedom.

For low energy electrons, the effective exchange constant $J$ is
renormalized due to the Kondo effect.\cite{Kondo} In the leading
logarithmic approximation \cite{Abrikosov} the renormalized exchange constant in Eq.~(\ref{KE}) is
\begin{equation}
J = \frac{2}{\nu} \ln^{-1}\displaystyle\frac{\varepsilon^*}{T_K},
\label{Jrg}
\end{equation}
where $\varepsilon^*$ is the characteristic energy of electrons
participating in the collision and $T_K$ is the Kondo
temperature. This approximation is justified as long as the energies
$\e_i\sim \e^*$of all incoming and outgoing electrons satisfy the
condition $\e^*\ ^>_\sim T_K$. It is important to note that energy
$\varepsilon^*$, which lies within the width of the electron
distribution function, does not cut off the singularity in the
transferred energy $E$. For a more detailed expression for the
renormalized $K(E)$ see Ref.~\cite{KG}.

The low-energy divergence of the inelastic scattering amplitude
(\ref{inamp}) is cut off by the time evolution of the impurity spin
correlator $\langle S'| \hat{S}^{j}(t) \hat{S}^{k}(t')| S\rangle$.  In
magnetic field $B$ this evolution is a spin precession with frequency
$\omega_{\rm s}=g_{\rm imp}\mu_{\rm B}B$.  When $\omega_{\rm s}$
exceeds the energies of the electrons being scattered, the scattering
rate saturates \cite{GGAG} at
\begin{equation}
  \label{KEZeeman}
K(E)\sim \frac{n}{\nu}S(S+1)(J\nu)^4
\frac{1}{\omega_{\rm s}^2}.
\end{equation}
The scattering processes in which both initial or both final electrons have
the same spin are suppressed completely.

The other mechanism, which cuts off the $E=0$ singularity of the
kernel (\ref{KE}) even at $B=0$, is the impurity spin relaxation. This
relaxation limits the lifetime of the intermediate state and the
denominator in Eq.~(\ref{inamp}) acquires the imaginary part.  At high
temperature $T>T_K$ scattering of the thermal electrons on the spin
results in an exponential decay of the spin correlation function,
$\langle S'| \hat{S}^{j}(t) \hat{S}^{k}(t')| S\rangle\propto
\exp(-|t-t'|/\tau_T)$.  The impurity spin correlation time $\tau_T$
can be evaluated with the help of the Fermi golden rule. If the
deviation from the thermal equilibrium is weak, we have
\begin{equation}
\frac{\hbar}{\tau_{T}}
=\frac{2\pi}{3}  (J\nu)^2 T\;.
\label{spin-flip-born}
\end{equation}
Here, as $T$ is lowered towards $T_K$, the exchange constant 
is renormalized according to Eq.~(\ref{Jrg}). 
The energy scale $\hbar/\tau_T$ sets the limit of applicability of
Eq.~(\ref{inamp}) and cuts off the singularity in the kernel (\ref{KE}) at
$E\sim \hbar/\tau_T$. Note that at $T>T_K$ the renormalized spin-flip rate satisfies the condition $\hbar/\tau_T>T_K$.

At very small energies ($|\varepsilon_i|,T\ll T_K$) the
Fermi-liquid description of electrons is valid again. The
behavior of the system is described in this case by the quadratic
fixed-point Hamiltonian, with the four-fermion interaction being the
least-irrelevant term.\cite{Nozieres,AffleckLudwig93} The calculation
of the inelastic scattering rate is then straightforward. The
resulting collision-integral kernel is energy-independent:
$K(E)=n/(\nu T_K^2)$.

We also discuss the relaxation due to the electron 
scattering on magnetic impurities in wires with applied bias $eV\gg T$.  
In this case the electron distribution is smeared, and the width of smearing
$eV$ exceeds the typical energies $|\varepsilon_i|$ of the
colliding electrons.  Assuming $eV\gg T_K$ and substituting the renormalized constant $J$, see Eq.~(\ref{Jrg}), 
into the kernel Eq.~(\ref{KE}), we obtain
\begin{equation}
  \label{KEV}
K(E)= \frac{\pi}{2}\frac{n}{\nu}S(S+1)[\ln(eV/T_K)]^{-4}
\frac{1}{E^2}\;.
\end{equation}
The $1/E^2$ dependence in Eq.~(\ref{KEV}) persists down to the
cut-off, which is determined by the spin-flip rate $1/\tau_{eV}$.  
For the spin-flip rate in the non-equilibrium situation the temperature $T$ in $\tau_T$ 
should be replaced by the electron distribution function smearing $eV$:
\begin{equation}
\label{spin-flipRG}
\frac{\hbar}{\tau_{eV}}=\gamma [\ln(eV/T_K)]^{-2} eV.
\end{equation}
Here the numerical constant $\gamma\sim 1$ depends on details
of the non-equilibrium electron distribution function. 

\section{Effect of spin scattering on electron interference phenomena}
Magnetic impurities provide mechanism not only for electron energy relaxation but also for electron phase relaxation, which suppresses the interference phenomena, such as weak localization and conductance fluctuations. 
Here we present our results \cite{VG} for metal wires with magnetic impurities, which can be partially polarized by an applied magnetic field.\cite{Bobkovetal}

The weak localization correction to the conductivity of a wire without
spin-orbit scattering in the conditions of strong Zeeman splitting of
the conduction electron states ($\e_{\rm Z}\tau_{\rm s}\gg 1$) and slow 
impurity spin relaxation $(\tau_T\gg \tau_{\rm s})$ is \cite{VG}
\begin{eqnarray}
\Delta\sigma & = & 
-\frac{e^2}{4\pi\hbar T}
\int\frac{d\e}{\cosh^2\e/2T}
\frac{\sqrt{D\tau_{\rm s}}}{\sqrt{P(\e) + B^2/B_{\rm c}^2}}.
\label{eq:22}
\end{eqnarray}
Here $1/\tau_{\rm s}=2\pi\nu n J^2S(S+1)$ is the scattering rate of
electrons on magnetic impurities in the absence of magnetic field. 
Function $P(\e)$ represents the probability of an electron spin flip in the
presence of magnetic field $B$:
\begin{equation}
P(\e)=1-\frac{\langle \hat S_z^2 \rangle + 
                    \langle \hat S_z\rangle\tanh(\e+\os)/2T}
{S(S+1)} 
\label{Fqs}
\end{equation}
For  $S=1/2$ impurities, we have $\langle \hat S_z^2 \rangle=1/4$ and
$\langle \hat S_z \rangle =(1/2)\tanh(\omega_{\rm s}/2T)$. 
In this case function $P(\e)$ can be rewritten in the form:
\begin{equation}
P(\varepsilon)= \frac{4}{3}\left(
p_\downarrow (1-n(\e+\os))+p_\uparrow n(\e+\os)
\right),
\label{eq:cse1:2}
\end{equation}
where $p_{\uparrow,\downarrow} = e^{\pm \os/2T}/(2\cosh\os/2T)$ is the
probability for the impurity spin to be parallel (antiparallel) to the
direction of the magnetic field and $n(\e)=(1+\exp(\e/T))^{-1}$ is the
electron occupation number with energy $\e$.

The term $B^2/B_{\rm c}^2$ in Eq.~(\ref{eq:22}) represents the
orbital effect of the applied magnetic field on conduction electrons;
$B_{\rm c}$ defines the characteristic value of the magnetic field,
which produces the orbital dephasing rate comparable with the spin
scattering rate:
\begin{equation}
\label{eq:19}
B_{\rm c}=\vartheta \frac{\Phi_0}{\sqrt{D\ts A_{\rm w}}}; \;\;\; \Phi_0=\frac{2\pi\hbar c}{e}.
\end{equation}
Here $A_{\rm w}$ is the wire cross-section area and $\vartheta$ is a
dimensionless factor depending on the wire geometry and the magnetic
field orientation. The expression in the denominator, $\sqrt{D\ts
A_{\rm w}}$, represents the effective area covered by an electron
trajectory between consequent spin flips.  The characteristic magnetic
field $B_{\rm c}$ gives an upper estimate on system temperature
$T_{\rm c}$, below which the effect of spin polarization prevails over 
the orbital effect of magnetic field:
\begin{equation}
\label{eq:20}
T_{\rm c}= S g_{\rm imp} \muB B_{\rm c}.
\end{equation}

If the orbital effect of the magnetic field is strong, we expand
Eq.~(\ref{eq:22}) in $B_{\rm c}/B$ and obtain:
\begin{equation}
\Delta\sigma=- \frac{e^2}{\pi\hbar}
\sqrt{D\tau_{\rm s}}
\left(\frac{B_{\rm c}}{B}-
\frac{2}{3}\frac{B_{\rm c}^3}{B^3}\frac{\os}{T}\sinh^{-1} \frac{\os}{T}\right).
\label{eq:wl1:2}
\end{equation}

Conductance fluctuations can be considered similarly to the evaluation
of $\Delta\sigma$. We concentrate here on the amplitude of the
Aharonov-Bohm ``$hc/e$'' oscillations.  Magnetic field applied through
the ring changes electron wave functions and, consequently, the
conductance of the ring of radius $r$.  The conductance statistics is
characterized by the correlation function:
\begin{equation}
\langle\langle
g_{\Phi}g_{\Phi+\Delta \Phi}\rangle\rangle=\alpha\frac{e^4}{\hbar^2}
\sum_{k=0}^{\infty}  R_k\cos 2\pi k \frac{\Delta \Phi}{\Phi_0},
\label{eq:30}
\end{equation}
where $\Phi=\pi r^2 B$ is the magnetic flux through the ring and $\alpha$ is a dimensionless geometry dependent factor.

In the high temperature case, $\tau_{\rm s} T\gg 1$, we find the amplitude of
oscillations of the conductance correlation function \cite{VG}:
\begin{eqnarray}
{\cal R}_k & = & \frac{D^{3/2}}{r^3T^2}\int 
\frac{e^{-2\pi kr\sqrt{ \Gamma(\e)/D}}}{\sqrt{\Gamma(\e)}}
\frac{d\e}{\cosh^4\e/2T},
\label{eq:R}
\end{eqnarray}
$$
\Gamma(\varepsilon)=\gd+\frac{1}{\tau_{\rm{s}}}\left(
1-\frac{\langle \hat S_z \rangle^2 + \langle \hat S_z\rangle\tanh(\e+\os)/2T}
{S(S+1)}\right)
\nonumber
$$
with $\gd$ being the dephasing rate due to mechanisms other than
magnetic impurity scattering.

\section{Comparison with experiments}
Relaxation of the electron energy distribution was investigated
experimentally in metallic wires of Cu and Au in
Refs.~\cite{PothierEtal97,PierreEtal00}. In these
experiments, a finite bias $V$ was applied to
the wire terminals. It was found that starting from
fairly small wire lengths, the electron distribution is smeared over
the range of energies $eV$, instead of having two distinct steps
created by the bias applied to the wire ends.  The observed electron
energy relaxation was attributed~\cite{PothierEtal97,PierreEtal00} to
electron-electron collisions.
The collision-integral kernel for $E<eV$ extracted from the
experiments has the form $K(E)=\hbar/(\tau_0 E^2)$, with a cut-off at
some low energy, which scales linearly with $eV$.\cite{private}  

Properties of these samples are compatible with the presence of iron
impurities with a concentration up to few tens of ppm.\cite{private}
The spin-flip rate, Eq.~(\ref{spin-flipRG}), is the low-energy cut-off
for the $1/E^2$ kernel dependence.  This cut-off is roughly
proportional to the applied voltage, in agreement with experimental
observations.\cite{private} We must note, however, that the lower
voltages used in experiment~\cite{PierreEtal00} are close to the Kondo
temperature, so the leading-logarithmic
approximation~\cite{Abrikosov,Anderson}, used in derivation of
Eqs.~(\ref{KEV}), (\ref{spin-flipRG}), may be insufficient.

Recent experiments~\cite{AnthoreEtal01} demonstrate that the
previously observed~\cite{PothierEtal97} electron energy relaxation in
thin wires is indeed suppressed by the applied magnetic field, see Eq.~(\ref{KEZeeman}), thus
supporting our hypothesis that the origin of the relaxation is the
inelastic scattering on the magnetic impurities.

\begin{figure}[t]
\begin{center}\leavevmode
\includegraphics[width=.8\linewidth]{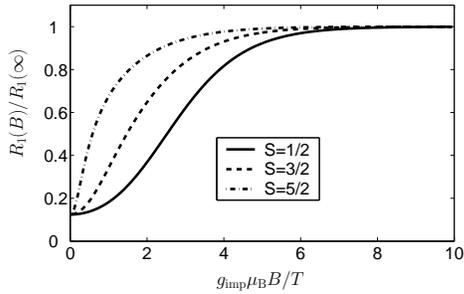}
\caption{ Figure shows dependence of the ``$hc/e$'' oscillations
of the conductance correlation function on the applied magnetic field $B$ for several values of the impurity spin $S$ in case when $\gd\tau_{\rm s}=1.5$.
 \label{fig2}}
\end{center}
\end{figure}

In measurements \cite{birge} of the conductance of a Cu ring, the
amplitude of the conductance oscillations increases in strong magnetic
field $\omega_{\rm c}\sim T$. This observation can be explained as the
result of the impurity spin polarization by the magnetic
field. Figure~\ref{fig2} represents the amplitude of the first
harmonic (``$hc/e$'' oscillations) of the conductance correlation function
in the limit $T\gg \gd,\ 1/\tau_{\rm s}$, described by
Eq.~(\ref{eq:R}), for different values of the impurity spin $S$.

In conclusion, the exchange interaction of itinerant electrons with
magnetic impurities can facilitate electron energy and phase
relaxation. We derived the kernel of the collision integral which
determines the energy relaxation, and found the weak localization
correction to the conductivity and the amplitude of conductance
fluctuations at an arbitrary level of polarization of magnetic
impurities by an external magnetic field. The obtained results provide
a quantitative explanation of the
experiments\cite{PothierEtal97,PierreEtal00} on anomalously strong
energy relaxation in short metallic wires and may be compared with the
observed behavior of the ``$hc/e$'' oscillations of the conductance
of an Aharonov-Bohm ring.\cite{birge}

\begin{ack}
The authors are grateful to I. Aleiner, N.
Birge,  and H. Pothier for valuable discussions.
This work was supported by NSF Grants No. DMR 97-31756 and DMR 0120702 at the University of Minnesota and by the
US-ONR, the LPS, and DARPA at the University of Maryland. 
\end{ack}

\end{document}